\documentclass{amsart}

\usepackage{epsfig}
\usepackage{amssymb}

\usepackage{amscd}
\usepackage{graphics}
\newtheorem{Theorem}{Theorem}[section]
\newtheorem{Proposition}{Proposition}[section]
\newtheorem{Corollary}{Corollary}[section]
\newtheorem{Lemma}{Lemma}[section]

\def\proof{\par{\it Proof}. \ignorespaces}
\def\endproof{{\hfill \vbox{\hrule\hbox{%
          \vrule height1.3ex\hskip0.8ex\vrule}\hrule }}\par}
\newenvironment{Proof}{\proof}{\endproof}

\def\sqr#1#2{{\vcenter{\vbox{\hrule height.#2pt
              \hbox{\vrule width.#2pt height#1pt \kern#1pt
                   \vrule width.#2pt}
            \hrule height.#2pt }}}}
\def\square{\mathchoice\sqr54\sqr54\sqr{2.1}3\sqr{1.5}3}

\theoremstyle{definition}
\newtheorem{Definition}[Theorem]{Definition}
\newtheorem{Example}[Theorem]{Example}

\theoremstyle{remark}
\newtheorem{Remark}[Theorem]{Remark}

\numberwithin{equation}{section}

\begin{document}

\renewcommand\baselinestretch{1.2}


\title{Young diagrams and $N$-soliton solutions of the KP equation}

\author{Yuji Kodama}
\address{Department of Mathematics, Ohio State University, Columbus,
OH 43210}
\email{kodama@math.ohio-state.edu}



\begin{abstract}
We consider $N$-soliton solutions of the KP equation,
\[
(-4u_t+u_{xxx}+6uu_x)_x+3u_{yy}=0\,.
\]
An $N$-soliton solution is a solution $u(x,y,t)$ which has the same set
of $N$ line soliton solutions in both asymptotics $y\to\infty$ and $y\to -\infty$.
The $N$-soliton solutions include all possible resonant interactions
among those line solitons. We then classify those $N$-soliton solutions
by defining a pair of $N$-numbers $({\bf n}^+,{\bf n}^-)$ with
${\bf n}^{\pm}=(n_1^{\pm},\ldots,n_N^{\pm}),~n_j^{\pm}\in\{1,\ldots,2N\}$, which labels 
$N$ line solitons in the solution.
The classification is related to the Schubert decomposition of the Grassmann manifolds Gr$(N,2N)$, where
the solution of the KP equation is defined as a torus orbit.
Then the interaction pattern of $N$-soliton solution
can be described by the pair of Young diagrams associated with
$({\bf n}^+,{\bf n}^-)$.
We also show that $N$-soliton solutions of the KdV equation
obtained by the constraint $\partial u/\partial y=0$ cannot have
resonant interaction.
\end{abstract}

\maketitle
\markboth{YUJI KODAMA}
         {$N$ soliton solutions of the KP equation}

\thispagestyle{empty}
\pagenumbering{roman}

\pagenumbering{arabic}
\setcounter{page}{1}

\section{Introduction}
In this paper, we consider a family of exact solutions of the KP equation,
\[
\frac{\partial}{\partial x}\left(-4\frac{\partial u}{\partial t}+
\frac{\partial^3 u}{\partial x^3}+6u\frac{\partial u}{\partial x}\right)+
3\frac{\partial^2 u}{\partial y^2}=0\,.
\]
The solution $u(x,y,t)$ is obtained from the $\tau$-function $\tau(x,y,t)$ as
\[
u(x,y,t)=2\frac{\partial^2}{\partial x^2}\log \tau(x,y,t)\,.
\]
It is well known that some solutions can be obtained by the Wronskian
form (for example see \cite{freeman:83}),
\begin{equation}
\label{tau}
\tau={\rm Wr}(f_1,\ldots, f_N):=\left|
\begin{matrix}
f_1^{(0)} & \cdots & f_N^{(0)} \\
\vdots    & \ddots & \vdots    \\
f_1^{(N-1)} & \cdots & f_N^{(N-1)}
\end{matrix}
\right|\,,
\end{equation}
where $f_i^{(n)}=\partial^n f_i/\partial x^n$, and $\{ \, f_i(x,y,t)\,|\,i=1,\ldots,N\,\}$ is a linearly independent set of $N$ solutions
of the equations,
\[
\frac{\partial f_i}{\partial y}=\frac{\partial^2 f_i}{\partial x^2}\,\quad\quad
\frac{\partial f_i}{\partial t}=\frac{\partial^3 f_i}{\partial x^3}\,.
\]
Throughout this paper, we assume $f_i(x,y,t)$ to have the form,
\begin{equation}
\label{f-function}
f_i=\sum_{j=1}^M a_{ij}\,e^{\theta_j}\,,\quad {\rm for}\quad i=1,\ldots,N,\quad
{\rm and}~~M>N\,,
\end{equation}
with some constants $a_{ij}$ which define the $N\times M$ matrix $A_{(N,M)}:=(a_{ij})$, and the phase functions $\theta_j$ are given by
\begin{equation}
\label{theta}
\theta_j(x,y,t)=-k_jx+k_j^2y-k_j^3t+\theta_j^0\,\quad {\rm for}\quad j=1\ldots,M\,.
\end{equation}
Here $k_j$ and $\theta_j^0$ are arbitrary constants, and throughout this paper we assume $k_j$ being ordered as
\[
k_1<k_2<\cdots <k_M\,.
\]
Then choosing particular forms for the matrix $A_{(N,M)}$, one can obtain
several exact solutions of the KP equation. 
As the simplest case with $N=1$ and $M=2$, i.e.
$\tau=f_1=a_{11}e^{\theta_1}+a_{12}e^{\theta_2}$ with $a_{11}a_{12}>0$, we have 1-soliton solution,
\begin{equation}
\label{1soliton}
u=2\frac{\partial^2}{\partial x^2}\log \tau=\frac{1}{2}(k_1-k_2)^2
{\rm sech}^2\frac{1}{2}(\theta_1-\theta_2)\,.
\end{equation}
Here we assumed $a_{11},\,a_{12}>0$ and absorbed them to the constants $\theta_j^0$, so that we have $a_{11}=a_{12}=1$. Note here that if $a_{11}a_{12}<0$, then the $\tau$-function has zeros and the solution
blows up at some points in the $x$-$y$ plane.
In the $x$-$y$ plane, 1-soliton solution describes a plane wave $u=\Phi(k_xx+k_yy-\omega t)$ having the wavenumber vector ${\bf k}=(k_x,k_y)$
and the frequency $\omega$,
\[
{\bf k}=(-k_1+k_2,\, k_1^2-k_2^2) \quad\quad \omega=k_1^3-k_2^3\,.
\]
Here $({\bf k},\omega)$ satisfies the dispersion relation, $4\omega k_x+k_x^4+3k_y^2=0$. We refer to the 1-soliton solution as a line soliton,
which is expressed by a line $\theta_1=\theta_2$ in the $x$-$y$ plane for a fixed $t$.
Then the slope of the line $c=dx/dy=-k_y/k_x$ represnts the velocity of the line
soliton in the $x$-direction with respect to $y$; that is, $c=0$ indicates the direction of the positive $y$-axis. As in the paper \cite{biondini:03}, we also call a line soliton as $[i,j]$-soliton, if the soliton is parametrized by the pair $(k_i,k_j)$. In the case of (\ref{1soliton}), the soliton is $[1,2]$-soliton
with the velocity $c=k_1+k_2$.
Then the ordinary $N$-soliton solution consisting of $N$ line solitons,
$[2j-1,2j]$-soliton with $j=1,\ldots,N$, 
 is obtained from the following matrix $A_{(N,M)}$ with $M=2N$ \cite{hirota:88},
\[
A_{(N,2N)}=\begin{pmatrix}
1    &  1   &  0  &  0  &  0  &  \cdots & \cdots  &  0  &  0  &  0  \\
0    &  0   &  1  &  1  &  0  &  \cdots & \cdots  &  0  &  0  &  0  \\
\vdots & \vdots & \cdots & \cdots & \cdots & \ddots &\ddots &\vdots
&\vdots &\vdots\\
0    &  0   & \cdots & \cdots & \cdots & \cdots &\cdots & 0 & 1 & 1
\end{pmatrix}\,,
\]
Note here that all nonzero entries are normalized to be one by choosing
appropriate constants $\theta_j^0$, and
any $N\times N$ minor of the matrix $A_{(N,2N)}$ is non-negative.
The fact that all $N\times N$ minors are non-negative (or non-positive) is sufficient for the solution to be non singular (see below).

Also in the paper \cite{biondini:03}, we found that if all the minors
are nonzero and have the same sign, the solution $u$ gives an $(M-N,N)$-soliton solution consisting of $M-N$ incoming line solitons as $y\to -\infty$ and
$N$ outgoing line solitons as $y\to \infty$. The set of the $\tau$-functions with different size $N$ then provide the solution of the Toda lattice. In particular, if $M=2N$,
we have $N$-soliton solution in the sence that the solution has the same set of 
$N$ line solitons in both asymptotics for $y\to\pm\infty$ (i.e. the sets of incoming and outgoing solitons are the same). However as mentioned in \cite{biondini:03}, this $N$-soliton solution is different from the ordinary $N$-soliton solution, and the interaction of any pair of line solitons is
in resonance. In this case, the matrix $A_{(N,2N)}$ can be written in the following row reduced echelon form (RREF),
\[
A_{(N,2N)}=\begin{pmatrix}
1   &   0  &  \cdots  & 0   &  * & * & \cdots &  *  \\
0   &   1  &  \cdots  & 0   &  * & * & \cdots &  *  \\
\vdots & \vdots  & \ddots & \vdots &\vdots & \vdots & \ddots & \vdots\\
0   &   0  & \cdots  &  1   & * & *  & \cdots &  *  
\end{pmatrix}
\]
where the entries marked by ``$*$'' are chosen so that all the $N\times N$ minors
are nonzero and positive (i.e. all the $*$'s of the last row should be 
all positive, and those in the second row from the bottom are negative and so on). Note here that those $*$'s are all nonzero. Then the line solitons
in $N$-soliton solution of the Toda lattice hierarchy are given by
$[j,N+j]$-soliton for $j=1,\ldots,N$.

We then expect that the general case of $N$-soliton solutions has a mixed pattern consisting of resonant and non-resonant interactions among those line solitons. In this paper, we classfy all the possible $N$-soliton solutions
obtaind by the $\tau$-function (\ref{tau}) with $f_j$'s in (\ref{f-function})
and $M=2N$. The classification is to determine all possible patterns of interactions, and it is done by constructing a specific form of the coefficient
matrix $A_N:=A_{(N,2N)}$ in (\ref{f-function}) for each class of $N$-soliton solutions. It turns out that a complete classification of $N$-soliton solutions
can be obtained by the complementary pair of ordered sets of $N$-numbers, ${\bf n}^+=(n_1^+,\ldots,n_N^+)$ and
${\bf n}^-=(n_1^-,\ldots,n_N^-)$, which are related to the dominant exponentials in the $\tau$-function for $x\to\pm \infty$ and each soliton is given by
 $[n_j^+,n_j^-]$-soliton. Notice that with the set $\{1,\ldots,2N\}$, there are $(2N-1)!!$ number of different sets of
 $N$ pairs. We then claim that each set of $N$ pairs corresponds to an $N$-soliton solution, and provide an explicit way to construct all those $N$-soliton solutions.
 
It is also well-known that a solution of the KP hierarchy is a GL$(\infty)$-orbit on infinite dimensional Grassmannian (Sato Grassmannian)
\cite{sato:81}.
Then the $N$-soliton solution can be identified as the $2N$-dimensional
torus orbit on the Grassmannian Gr$(N,2N)$, and the coefficient matrix $A_N$
represnts a point on Gr$(N,2N)$. Since the Grassmannian has the Schubert cell decomposition, one can first classify the matrix $A_N$ using the decomposition,
i.e. identify the cell $A_N$ belongs to. This classification corresponds to the  set ${\bf n}^+$, and the ${\bf n}^-$ gives a further decomposition of the cells to classify the interaction patterns of $N$-soliton solutions. 
We then define a pair of Young diagrams $({ Y}^+,{Y}^-)$ associated with
the pair of the number sets $({\bf n}^+,{\bf n}^-)$.
In particular, the number of resonant vertices in the $N$-soliton solution
parametrized by $({Y}^+,{Y}^-)$ is given by
\[
\frac{N(N-1)}{2}-\left|Y^+\right|-\left|Y^-\right|\,,
\]
where $\left|Y^{\pm}\right|$ denote the size (degree) of the diagrams, i.e.
the number of boxes in the diagrams (Theorem \ref{num-res}).

The paper is organized as follows:
In Section 2, we give a general basic structure of the $\tau$-function, and present all types of $2$-soliton solutions, which provide the building blocks for the general case of $N$-soliton solutions. Here we also introduce the pair
of numbers $({\bf n}^+,{\bf n}^-)$, and the Young diagrams $(Y^+,Y^-)$.
In Section 3, we briefly introduce the Grassmannian Gr$(N,M)$ and
 the Schubert decomposition of Gr$(N,M)$. We also identify $N$-soliton solution as an 2$N$-dimensional
torus orbit on Gr$(N,2N)$, and briefly mention that Gr$(N,2N)$ contains all possible $(m,n)$-soliton solutions for $1\le m\le N$ and $1\le n \le N$, which
are distinguished by the Schubert decomposition. Here $(m,n)$-soliton is the solution consisting of $m$ incoming solitons 
for $y\to -\infty$ and $n$ outgoing solitons for $y\to\infty$ (see
\cite{biondini:03}). 
In Section 4, we describe the structure of the coefficient matrix $A_N$ for
each $N$-soliton solution by prescribing an explicit construction
of the matrix $A_N$. We here define {\it $N$-soliton condition}
on the matrix $A_N$ which determines local structures based on the types of
2-soliton interactions in the $N$-soliton solution. Finally, in Section \ref{KdV}, we discuss $N$-soliton solutions of the KdV equation, and show
that the KdV $N$-soliton solution cannot have resonant interactions.

\section{Basic structure of the $\tau$-function and 2-soliton solutions}
Let us start with the follwoing Lemma which shows the basic structure
of the $\tau$-function:
\begin{Lemma}
\label{binet}
The $\tau$-function (\ref{tau}) with $f_j$'s given in (\ref{f-function})
can be expanded as a sum of exponential functions,
\[
\tau=\sum_{1\le i_1<\cdots<i_N\le M}\xi(i_1,\ldots,i_N)\prod_{1\le j<l\le N}
(k_{i_j}-k_{i_l})\exp\left(\sum_{j=1}^{N}\theta_{i_j}\right)\,,
\]
where $\xi(i_1,\ldots,i_N)$ is the $N\times N$ minor given by the $i_j$-th columns with
$j=1,\ldots,N$ in the matrix $A_{(N,M)}=(a_{ij})$ of (\ref{f-function}),
\[
\xi(i_1,\ldots,i_N):=
\left|\begin{matrix}
a_{1,i_1}&\cdots& a_{1,i_N}\\
\vdots &\ddots &\vdots \\
a_{N,i_1}&\cdots&a_{N,i_N}
\end{matrix}
\right|\,.
\]
\end{Lemma}
\begin{Proof}
Apply the Binet-Cauchy theorem for the expression (see for example \cite{gantmacher:59}),
\[
\tau=\left|\begin{pmatrix}
E^{(0)}_1 & E^{(0)}_2 & \cdots & E^{(0)}_M\\
E^{(1)}_1 & E^{(1)}_2 & \cdots & E^{(1)}_M\\
\vdots    &  \vdots   & \ddots & \vdots  \\
E^{(N-1)}_1 & E^{(N-1)}_2 & \cdots & E^{(N-1)}_M
\end{pmatrix}
\begin{pmatrix}
a_{11}  &  a_{21} &  \cdots  & a_{N1}\\
a_{12}  &  a_{22} &  \cdots  & a_{N2}\\
\vdots  &  \vdots & \ddots   & \vdots\\
\vdots  & \vdots  & \ddots   & \vdots\\
a_{1M}  &  a_{2M} &  \cdots  & a_{NM}
\end{pmatrix}
\right|\,,
\]
where $E^{(n)}_j=(-k_j)^nE_j$ with $E_j=e^{\theta_j}$.
\end{Proof}
 From this Lemma, it is clear that if all the
 minors $\xi(i_1,\ldots,i_N)$ are non-negative (or non-positive), then
 the $\tau$-function is sign definite and the corresponding solution $u$ is nonsingular (recall also the order $k_1<\cdots<k_M$). This is the first restriction on the matrix $A_{(N,M)}$
for $N$-soliton solution, and
we later impose that
all the $N\times N$ minors are non-negative after making $A_{(N,M)}$ to be
in the row reduced echelon form.

We now consider the case of $N$-soliton solutions, that is, we take $M=2N$.
With the ordering of the numbers $k_j$'s, i.e. $k_1<k_2<\cdots<k_{2N}$,
one can find the asymptotic behavious of the $\tau$-function for $x\to\pm\infty$: For this purpose, we first define the function $w$ given by
\[
w(x,y,t)=-\frac{\partial}{\partial x}\log \tau(x,y,t)\,.
\]
Suppose that $w$ has the asymptotic form with some numbers $\{n_1^{\pm},\ldots,n_N^{\pm}\}\subset\{1,\ldots,2N\}$,
\begin{equation}
\label{asymptotics}
w\longrightarrow \left\{
\begin{array}{ccccc}
\displaystyle{\sum_{j=1}^N k_{n_j^+}\,} \quad &{\rm as}\quad & x\to \infty\,, \\
{}\\
\displaystyle{\sum_{j=1}^Nk_{n_j^-}\,}\quad & {\rm as}\quad & x\to -\infty\,.
\end{array}
\right.
\end{equation}
It is clear that $n_i^+\ne n_j^+$ and $n_i^-\ne n_j^-$ if $i\ne j$.
Then with the sets of $N$ numbers $n_j^{\pm}$'s in the asymptotics (\ref{asymptotics}), we define the following two $N$-vectors with the entries
 from the set $\{1,2,\ldots,2N\}$:
\begin{Definition}
A pair $({\bf n}^+,{\bf n}^-)$ of $N$-vectors are defined by
\[
\left\{
\begin{array}{lll}
{\bf n}^+ =(n_1^+,n_2^+,\ldots,n_N^+)\quad {\rm with}\quad 1=n_1^+<n_2^+<\cdots<n_N^+<2N,\\
{}\\
{\bf n}^- =(n_1^-, n_2^-, \ldots, n_N^-)\quad {\rm with}\quad n_j^->n_j^+\,.
\end{array}\right.
\]
The ordering in the set ${\bf n}^+$ is just a convenience, and once we made
the ordering in ${\bf n}^+$, we do not assume a further ordering for other set ${\bf n}^-$. 
We assume that those sets are complementary in the set $\{1,2,\ldots,2N\}$. As will be explained below,
this assumption is necessary to have an $N$-soliton solution, that is, the solution
has the same sets of $N$ line solitons for both asymptotics $y\to\pm\infty$.
\end{Definition}

With those number sets ${\bf n}^+$ and ${\bf n}^-$, one can define the
pair of Young diagrams $(Y^+,Y^-)$:
\begin{Definition}
\label{young}
We define a pair $(Y^+,Y^-)$ associated to the pair $({\bf n}^+,{\bf n}^-)$:
\begin{itemize}
\item{} The $Y^+$ is a Young diagram represented by $(\ell_1^+,\ldots,
\ell_N^+)$ with $\ell_j^+=n_j^+-j$ where each $\ell_j^+$ represents the number of boxes in a row counting from the bottom (note $\ell_j^+\le\ell^+_{j+1}$).
\item{} The $Y^-$ is a Young diagram associated to $(\ell_1^-,\ldots,\ell_N^-)$, where $\ell_j^-$ is defined as
\[
\ell_j^-=\left|\left\{\, n_k^-\,\Big|\, n_k^-<n_j^-\,~{\rm for}~\,k>j\,\right\}\right|\,.
\]
Then $\ell_j^-$ represents the number of boxes in a row arranged in
the increasing order from the bottom.
\end{itemize}
\end{Definition}
Notice that the Young diagram $Y^+$ is uniquely determined from ${\bf n}^+$, but
the diagram $Y^-$ is not unique (note that $(\ell_1^-,\ldots,\ell_N^-)$ is unique of course). However the number of boxes in $Y^-$
gives the number of intersection vertices of $N$-soliton solution
having a particular type of interaction (Theorem \ref{num-res}).

As a main goal of this paper, we will show in Section \ref{Amatrix} that for each pair $({\bf n}^+,{\bf n}^-)$
one can construct an $N$-soliton solution consisting of $N$ line solitons, each of which 
is given by $[n_j^+,n_j^-]$-soliton for $j=1,\ldots,N$.
The total number of different $N$-soliton solutions is given by $(2N-1)!!$ which is
the number of different sets of pairs $\{\,({\bf n}^+,{\bf n}^-)\,\}$. 
Namely, we construct an $N$-soliton solution which is labeled by $({\bf n}^+,{\bf n}^-)$ as follows: 
\begin{itemize}
\item[1)] Consider a set of $2N$ numbers $\{1,\ldots,2N\}$, which are associated to the parameters $k_1<\cdots<k_{2N}$.
\item[2)] Choose $N$ different pairs from $\{1,\ldots,2N\}$, and label each
pair as $[n_j^+,n_j^-]$ so that the numbers $n_j^{\pm}$ satisfy
\begin{itemize}
\item[i)] $n_j^+<n_j^-$ for $j=1,\ldots, N$,
\item[ii)] $1=n_1^+<n_2^+<\cdots<n_N^+$.
\end{itemize}
\item[3)] Define the pair $({\bf n}^+,{\bf n}^-)$ with ${\bf n}^{\pm}=(n_1^{\pm},\ldots,n_N^{\pm})$, and construct the corresponding
matrix $A_N=A_{(N,2N)}$.
\end{itemize}
Note here that the labeling is unique with a given choice of pairs, and the corresponding $N$-soliton solution consists of $[n_j^+,n_j^-]$-solitons for $j=1,\ldots,N$. The item 3) will be given in Section \ref{Amatrix}.

In order to classify those $N$-soliton solutions,  
we first note: 
\begin{Lemma}
\label{numbernminus}
For each given ordered set ${\bf n}^+$,
the number of choices of ${\bf n}^-$ is given by
\[
m_{{\bf n}^+}=\prod_{j=1}^N (2j-n_j^+)\,.
\]
\end{Lemma}
\begin{Proof}
Since $n_N^->n_N^+$, there are $(2N-n_N^+)$ numbers of possible choices of $n_N^-$.
Having made a choice of $n_N^-$, one has $2(N-1)-n_{N-1}^+$ many choices for
$n_{N-1}^-$. Now repeating this, the result is obvious.
\end{Proof}
This Lemma provides the number of different $N$-soliton solutions
having the same ${\bf n}^+$, that is, the functions
$w=-(\partial/\partial x)\log\tau$ for those solutions have the same asymptotic values for $x\to\pm\infty$.
As a corollary of this Lemma, we also have the following identity,
\[
(2N-1)!!=\sum_{{\bf n}^+}m_{{\bf n}^+}\,.
\]

We also have the following Lemma for the ordered set ${\bf n}^+$:
\begin{Lemma}
Each number $n_j^+$ in the ${\bf n}^+$ is limited as
\[
n_j^+\le 2j-1\, \quad {\rm for}\quad j=1,\ldots, N\,.
\]
\end{Lemma}
\begin{Proof}
Since $n_j^+<n_j^-$, there are $N-j$ numbers of $n_k^+$ and $N-j+1$
of $n_k^-$, which are larger than $n_j^+$, that is, we have
\[
(N-j)+(N-j+1)\le 2N-n_j^+\,.
\]
This implies the Lemma.
\end{Proof}

In terms of the Young diagram $Y^+$, we have
\begin{Corollary}
The maximum diagram for $Y^+$, denoted as $Y_{\rm max}$, is the Young diagram associated to ${\bf n}^+=(1,3,\ldots,2N-1)$, that is,
$Y_{\rm max}$ is the up side down staircase with the size,
\[
\left|Y_{\rm max}\right|=\frac{N(N-1)}{2}\,.
\]
The maximum Young diagram for $Y^-$ is also given by $Y_{\rm max}$,
which now corresponds to the case ${\bf n}^-=(2N,2N-1,\ldots,N+1)$.
Then the Young diagrams $Y^{\pm}$ associated to $N$-soliton solutions are given by
a subdiagram of $Y_{\rm max}$.
\end{Corollary}
Note that $|Y_{\rm max}|$ is also the number of intersection vertices of $N$ lines in a plane
in the general position. We will then determine whether each point is of
 resonant or non-resonant.
 
\begin{figure}[t!]
\epsfig{file=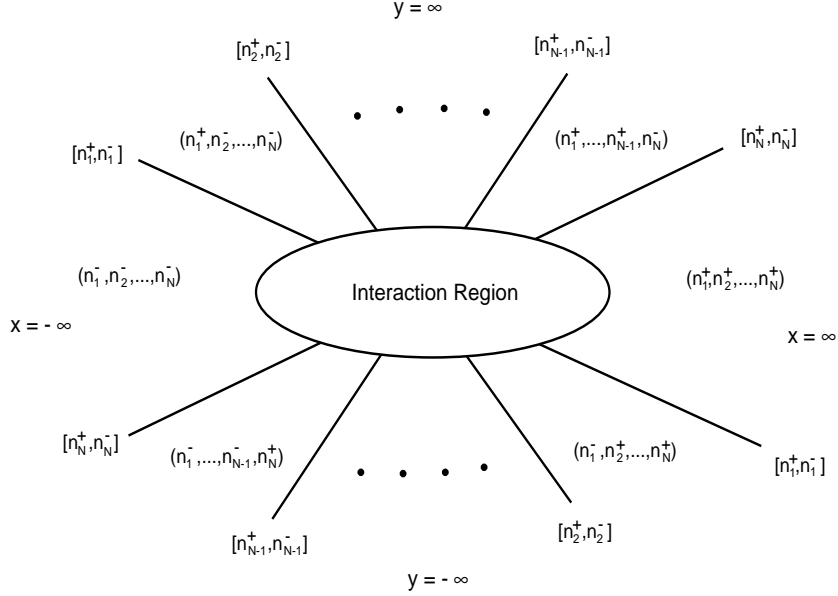,height=8cm,width=11cm}
\caption{An $N$-soliton solution. The label $[n_j^+,n_j^-]$ indicates the $[n_j^+,n_j^-]$-soliton. Here the soliton velocities 
$c_j:=k_{n_j^+}+k_{n_j^-}$ are assumed to be ordered as $c_{j+1}>c_{j}$.
Each asymptotic region is marked by the value of the function $w$ there, i.e. $(i_1,\ldots,i_N)$ implies $w=\sum_{j=1}^N k_{i_j}$.}
\label{Nsol:fig}
\end{figure}

In Figure \ref{Nsol:fig}, we illustrate the asymptotic stage of an $N$-soliton solution with $N$ solitons marked by $[n_j^+,n_j^-]$ for $j=1,\ldots,N$.
We note a duality of the values of $w$ in the sense that if $w$ takes
$\sum_{j=1}^Nk_{i_j}$, denoted as $(i_1,\ldots,i_N)$ in the Figure,
then $w$ also takes $(j_1,\ldots,j_N)$ which is the complementary set of
$(i_1,\ldots,i_N)$ in $\{1,\ldots, 2N\}$. In terms of the minors, this implies
that if $\xi(i_1,\ldots,i_N)\ne 0$ then $\xi(j_1,\ldots,j_N)\ne 0$.
The duality can be considered as a symmetry of the KP equation:
In particular, if $\theta_j^0=0$ for all $j$, then $N$-soliton solution
has the symmetry $(x,y,t)\leftrightarrow (-x,-y,-t)$.
In this case, at $t=0$ all the solitons intersect at the origin in the
$x$-$y$ plane (see Figure \ref{4sol:fig}). Then each soliton can be
moved by shifting $\theta_j^0$.
With this symmetry, we define the duality of the $\tau$-function:
\begin{Definition}
\label{duality}
We say that
the $\tau$-function in Lemma \ref{binet} with $M=2N$ satisfies {\it the duality},
if for any complementary sets $\{i_1,\ldots,i_N\}$ and $\{j_1,\ldots,j_N\}$,
the minors satisfy
\[
\xi(i_1,\ldots,i_N)=0\quad {\rm if~and~only~if}\quad \xi(j_1,\ldots,j_N)=0\,.
\]
\end{Definition}

\medskip

Before we discuss the general case with arbitrary $N$, we list up all possible 2-soliton solutions, and show that the $\tau$-functions
for those cases all satisfy the duality. As will be shown below, those
solutions give the building blocks of $N$-soliton solutions.
There are three cases of the solutions (see also \cite{infeld:90}),
which are labeled by the pair $({\bf n}^+,{\bf n}^-)$ with ${\bf n}^{\pm}=
(n_1^{\pm},n_2^{\pm})$. Two line solitons are labeled by $[n_1^+,n_1^-]$
and $[n_2^+,n_2^-]$:
\begin{itemize}

\item[i)] ${\bf n}^+=(1,3)$ and ${\bf n}^-=(2,4)$: This corresponds to the ordinary 2-soliton solution, and the matrix $A_2$ which we denote as $A^{(O)}_2$, takes
the form (after choosing $\theta_j^0$ properly),
\[
A^{(O)}_2=\begin{pmatrix}
1  &  1  &  0  &  0  \\
0  &  0  &  1  &  1  
\end{pmatrix}
\]
which gives only 4 nonzero minors, $\xi(1,3),\,\xi(1,4),\,\xi(2,3)$ and $\xi(2,4)$. This shows the duality, i.e. $\xi(1,2)=\xi(3,4)=0$.
Those nonzero minors then label the four asymptotic
regions separated by two line solitons:  Denoting $(i,j):=k_i+k_j$, $w$ takes $(1,3)$ as $x\to\infty$
which is dual to $(2,4)$ as $x\to -\infty$. Also $w$ takes $(1,4)$ as
$y\to\infty$, which is dual to $(2,3)$ as $y\to -\infty$. 
 This also implies that there is no other transition
of the $w$ value, hence the interaction is not a resonant type.

The Young diagram $Y^+$ in this case is the maximum one, i.e. $Y^+=\square$. The diagram $Y^-$ has no box, $Y^-=\emptyset$.
The interaction of O-type is thus labeled by the Youg diagram $Y^+$. 
It is also useful to note that the intervals defined by the labels $[1,2]$
and $[3,4]$ of solitons have no overlaps. Non-overlapping property can be 
stated as $\xi(1,2)=0$ and its dual one $\xi(3,4)=0$.

\item[ii)] ${\bf n}^+=(1,2)$ and ${\bf n}^-=(3,4)$: This case corresponds to
the 2-soliton solution of the Toda lattice \cite{biondini:03}, and the matrix
$A_2$, denoted as $A_2^{(T)}$, has the structure,
\[
A^{(T)}_2=\begin{pmatrix}
1  &  0  &  -  &  -  \\
0  & 1  &  +   &   +  
\end{pmatrix}\,,
\]
where ``$+,\,-$'' shows the signs of the entries (also nonzero).
An explicit example of $A_2^{(T)}$ is
\[
A^{(T)}_2=\begin{pmatrix}
1  &  0  &  -1 &  -2  \\
0  & 1  &  1  &   1  
\end{pmatrix}\,.
\]
and the corresponding 2-soliton solution is illustrated as T-type in Figure \ref{2sol:fig}.
 Following the arguments in \cite{biondini:03}, one
can easily show that two solitons are given by $[1,3]$- and $[2,4]$-solitons, and the interaction is in resonance. The main point in this example is that
all the $2\times 2$ minors of $A_2^{(T)}$ are nonzero, and this results resonance interaction. Non-vanishing condition of all the minors is necessary to make the resonsnt interaction, and those minors describe the intermediate
solitons which forms $Y$-shape resonant interaction (see \cite{biondini:03}
for more detail). 

Again we notice the duality in the values of $w$
in the asymptotics: $w$ takes $(3,4)$ for $x\to -\infty$
which is dual to $(1,2)$ for $x\to\infty$. Also $w$ takes $(1,4)$ for $y\to\infty$, which is dual to $(2,3)$ for $y\to -\infty$. One more duality
appears in the inside of the resonant hole, that is, $w$ takes $(1,3)$
for $t>0$, which is dual to $(2,4)$ for $t<0$ (or vice versa).
Thus all six nonzero minors contribute to make the resonant interaction.
The pair $(Y^+,Y^-)$ of the Young diagrams is then given by $(\emptyset,\,\emptyset)$. Alos note that the intervals defined by
the labels of solitons, $[1,3]$ and $[2,4]$ have a partial overlap.

\begin{figure}[t!]
\epsfig{file=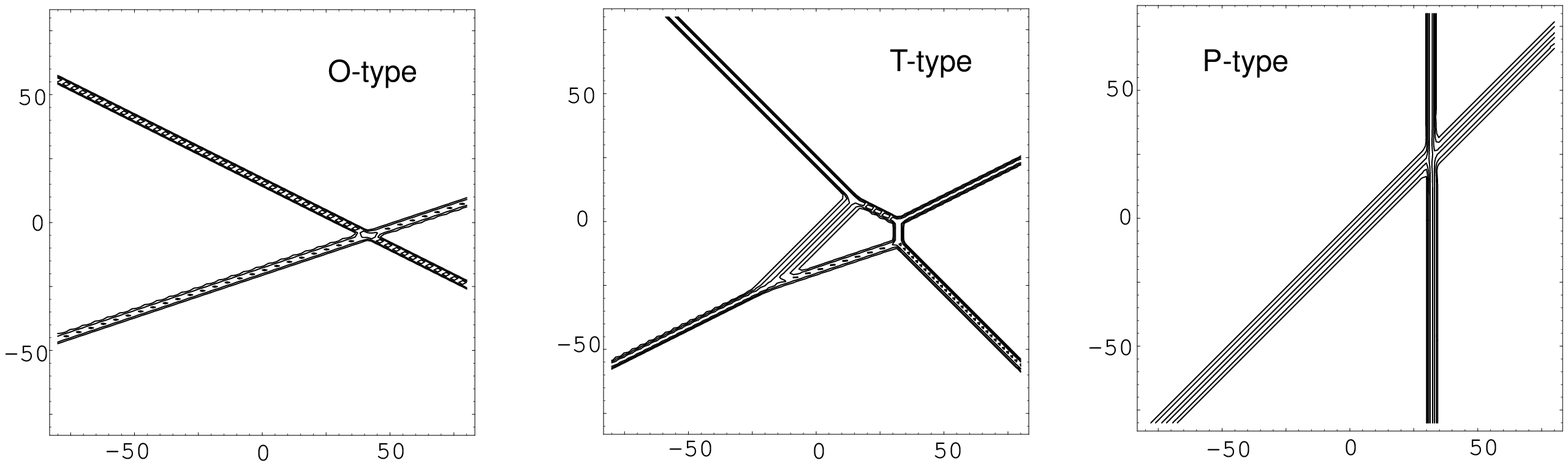,height=3.5cm,width=12cm}
\caption{2-soliton solutions. Two line solitons in
those figures are $[1,2]$- and $[3,4]$-solitons for
O-type, $[1,3]$- and $[2,4]$-solitons for T-type,
and $[1,4]$-and $[2,3]$-solitons for P-type. The parameters are chosen as
$(k_1,k_2,k_3,k_4)=(-2,0,1,2)$ and $\theta_j^0=0,\,\forall j$ for
all types.}
\label{2sol:fig}
\end{figure}

\item[iii)] ${\bf n}^+=(1,2)$ and ${\bf n}^-=(4,3)$: This case has been noted in
\cite{infeld:90}, and the corresponding matrix $A_2$, denoted as $A_2^{(P)}$
(``$P$'' stands for {\it Physical}, see Remark \ref{physical}), is given by
(after choosing $\theta_j^0$ properly),
\[
A_2^{(P)}=\begin{pmatrix}
1  &  0  &  0  &  -1 \\
0  &  1  &  1  &  0  
\end{pmatrix}\,,
\]
which gives again only four nonzero minors with $\xi(1,2)$, $\xi(1,3)$, $\xi(2,4)$ and
$\xi(3,4)$. This implies also no resonance, and the solitons are given by
$[1,4]$- and $[2,3]$-solitons. This solution is similar to the O-type, but the pair of the Young diagrams is different, and given by $(\emptyset, \square)$.
This interaction of this type is labeled by the Young diagram $Y^-$.
Also the intervals defined by the labels of solitons, $[1,4]$ and $[2,3]$,
have complete overlap. This implies $\xi(2,3)=0$ and its dual $\xi(1,4)=0$.
\end{itemize}

One should note that we have the following order in the soliton velocities 
$c_{ij}:=k_i+k_j$ among those solitons for given constants $k_j$ with the order $k_1<k_2<k_3<k_4$,
\[
c_{12}<c_{13}<c_{14},\, c_{23} <c_{24}<c_{34}\,.
\]
Here note that $c_{14}$ and $c_{23}$ cannot have a definite order (even $c_{14}=c_{23}$ is possible).
This ordering indicates that any 2-soliton solution can be classified as one of
those three solitons. One should note here that the type for a given pair of
solitons are completely determined by their labels $[n_j^+,n_j^-]$ for $j=1,2$. Namely the inetraction pattern is completely determined their
asymptotics in $y\to\infty$, where two solitons are labeled.

Although the interaction patterns of the cases of O-type in ii) and P-type in iii) are the same, i.e. nonresonant, they describe orbits
on different cells in terms of the Schubert decomposition of the Grassmannian,
in which each cell is parametrized by a Young diagram.
We will explain more details on this description in Section \ref{grassmann}.

\begin{Remark}\label{physical}
In \cite{kp:70}, the KP equation was introduced to describe a transversal stability of the KdV soliton propagating along the $x$-axis. Then the order
of the wavenumber in the $y$-direction is assumed to be much smaller
than that in the $x$-direction. This implies that the wave described by the KP
equation should be almost parallel to the $y$-axis for a better approximation
within the physical setting. In this sense, the P-type solitons
are more physical than other two types. The KdV two solitons are also obtained from the P-type solitons with $k_1<k_2<0$ and $k_3=-k_2$, $k_4=-k_1$ (see
Section \ref{KdV}).
\end{Remark}

\section{Grassmannian Gr$(N,2N)$ and $N$-soliton solutions}
\label{grassmann}

Here we briefly summarize the basics of the Grassmann manifold Gr$(N,M)$
in order to explain that the minors $\xi(i_1,\ldots,i_N)$ in Lemma \ref{binet} provide
a coordinate system for Gr$(N,M)$, the Pl\"ucker coordinates.
Namely a solution given by the $\tau$-function (\ref{tau})
can be marked by a point on Gr$(N,M)$.
The purpose of this section is to identify an $N$-soliton solution as
a $2N$-dimensional torus orbit of a point on Gr$(N,2N)$ marked by
the minors $\xi(i_1,\ldots,i_N)$.
Then using the Schubert decomposition of the Grassmannian Gr$(N,2N)$,
we classify the orbits which represent the $N$-soliton solutions.

\subsection{Grassmannian Gr$(N,M)$}
A real Grassmannian Gr$(N,M)$ is the set of $N$-dimensional subspaces
of ${\mathbb R}^{M}$. A point $\xi$ of the Grassmannian is expressed by
the $N$-frame of vectors,
\[\xi=[\xi_1,\xi_2,\cdots,\xi_N], \quad {\rm with}\quad
\xi_i=\sum_{j=1}^Ma_{ij}e_j\in {\mathbb R}^{M}\,,\]
where $\{e_i~|~i=1,2,\cdots,M\}$ is the standard basis of
$\mathbb R^{M}$, and $(a_{ij})=A_{(N,M)}$ is the $N\times M$
matrix given in (\ref{f-function}).
Then the Grassmannian Gr$(N,M)$ can be embedded to the
projectivization of the exterior space $\bigwedge^{N}{\mathbb
R}^{M}$, which is called the Pl\"ucker embedding,
\[
\begin{matrix}
{\rm Gr}(N,M) &\hookrightarrow &{\mathbb P}(\bigwedge^{N}{\mathbb R}^{M})\\
\xi=[\xi_1,\cdots,\xi_{N}] &\mapsto & \xi_1\wedge\cdots\wedge\xi_{N}
\end{matrix}
\]
Here the element on $\mathbb P(\bigwedge^{N}{\mathbb R}^{M})$ is
expressed as
\[
\xi_1\wedge\cdots\wedge\xi_N=\sum_{1\le i_1<\cdots<i_N\le
M}\xi{(i_1,\cdots,i_N)}\,e_{i_1}\wedge\cdots\wedge e_{i_N}\,,
\]
where the coefficients $\xi{(i_1,\cdots,i_N)}$ is an $N\times N$ minors
given in Lemma \ref{binet}, which are called the Pl\"ucker
coordinates.

It is also well known that the Grassmannian can have the cellular 
decomposition, called the Schubert decomposition \cite{griffiths:78},
\begin{equation}
\label{Wcell}
{\rm Gr}(N,M)=\bigsqcup_{1\le i_1<\cdots<i_N\le M} W{(i_1,\cdots,i_N)}
\end{equation}
where the cells are defined by
\[\begin{array}{lll}
W{(i_1,\cdots,i_N)}&=&\left\{~\xi=
\begin{pmatrix}
0    & 0   &  0    & \cdots & 0 \\
\vdots &\vdots & \vdots & \vdots &\vdots \\
0    & 0   &  0    &  \cdots  & 0\\
1    & 0   & 0     & \cdots   & 0\\
*    & 0   & 0     & \cdots   & 0\\
0    & 1   & 0     & \cdots   & 0\\
*    & *   &  \cdots &\cdots  & 0\\
0    & 0   & 0     &\cdots   & 1\\
\vdots& \vdots &\vdots &\vdots& \vdots\\
*    & *   &  *   & \cdots    & *
\end{pmatrix}
 ~\in ~\overbrace{ \phantom{\bigg|}\kern-0.2em
{\mathbb R}^{M}\times\cdots\times{\mathbb R}^{M}}^{N}~
\right\}\\
{}\\
&=&\{~{\rm the~matrix~}\xi=A_{(N,M)}^T~{\rm in~the~
echelon~form~}\\
&{}& {\rm whose~pivot~ones~are~at~}(i_1,\cdots,i_N)~{\rm positions}~\}
\end{array}
\]
Namely an element $\xi=[\xi_1,\cdots,\xi_N]\in W{(i_1,\cdots,i_N)}$ is
described by
\[
\xi\in W{(i_1,\cdots,i_N)}~\Leftrightarrow~\left\{
\begin{array}{llll}
{\rm (i)}&
\xi{(i_1,\cdots,i_N)}\ne 0.\\
 {\rm (ii)}&
\xi{(j_1,\cdots,j_N)}=0~{\rm if}~ j_n<i_n~{\rm  for~some}~ n.
\end{array}
\right.
\]
Each cell $W{(i_1,\cdots,i_N)}$ is called the Schubert cell, and it is convenient to label the cell by the Young diagram
$Y=(i_1,\cdots,i_k)$ where the number of boxes are given by $\ell_j=i_j-j$
for $j=1,\cdots,N$ (counted from the bottom), which also expresses a partition
$(\ell_N,\ell_{N-1},\cdots,\ell_1)$ of the number $|Y|:=\sum_{i=1}^N\ell_i$,
the {\it size} of $Y$. 
The codimension of the cell $W{(i_1,\cdots,i_k)}$ is then given by the
size of $Y$, i.e.
\[
{\rm codim}\,W{(i_1,\cdots,i_k)}=|Y|,
\]
and the dimension is given by the number of free variables in the echelon form.
Note that the top cell of Gr$(N,M)$ is labeled by $Y=(1,\cdots,N)$, i.e.
$|Y|=0$, and
\[
{\rm dim}~W{(1,\cdots,N)}={\rm dim}~{\rm Gr}(N,M)=N\times (M-N).
\]

The main point here is that each matrix $A_{(N,M)}$ in (\ref{f-function})
can be identified as a point on a Schubert cell of the Grassmannian.
Now we identify an $N$-soliton solution as a torus orbit of a point on the Grassmannian Gr$(N,2N)$, and show that orbits in the different cells
give different types of $N$-soliton solutions. 

\subsection{$N$-soliton solution as a $2N$-dimensional torus orbit}
First we note that the $\tau$-function (\ref{tau}) in the Wronskian form
can be expressed as
\begin{equation}
\label{orbit}
\tau(x,y,t)=\langle \, v^N,\,KD(x,y,t)A^t\cdot v^N\,\rangle\,,
\end{equation}
where $v^N$ is the highest weight vector in a fundamental representation of GL$(2N)$ which is given by an element of $\bigwedge^N{\mathbb R}^{2N}$,
\[
v^N=e_1\wedge e_2\wedge \cdots\wedge e_N\,, 
\]
and $K,\,A$ ($A^t$ represents the transpose of $A$) are $2N\times 2N$ constant matrices defined by
\[\left\{
\begin{array}{lllll}
&\displaystyle{K=(k_{ij})_{1\le i,j\le 2N}\,, \quad {\rm with}\quad k_{ij}:=(-k_j)^{i-1}}\\
&{}\\
&\displaystyle{A=(a_{ij})_{1\le i,j \le 2N}}\,.
\end{array}\right.
\]
The $D(x,y,t)$ is $2N\times 2N$ diagonal matrix associated to an $({\mathbb R}^*)^{2N}$-torus action,
\[
\displaystyle{D(x,y,t)={\rm diag}(e^{\theta_1},\ldots,e^{\theta_{2N}})}
\quad{\rm with}\quad \theta_j=-k_jx+k_j^2y-k_j^3t+\theta_j^0\,.
\]
Note here that $A^t\cdot v^N$ defines the matrix $A_N=A_{(N,2N)}$ in (\ref{f-function}), that is,
\[
A^t\cdot e_1\wedge\cdots\wedge e_N=A^t e_1\wedge A^t e_2\wedge\ldots A^t e_N\,,
\]
which can be idetified as $A_N^t$. The inner product $\langle \cdot,\cdot\rangle$
on $\bigwedge^N{\mathbb R}^{2N}$ is defined by
\[
\langle v_1\wedge\cdots\wedge v_j,~w_1\wedge\cdots\wedge w_j\rangle
={\rm det}\left[(\langle v_m,w_n\rangle)_{1\le m,n\le j}\right]\,,
\]
where $\langle v_m,w_n\rangle$ is the standard inner product of $v_m,\,w_n
\in {\mathbb R}^{2N}$.

 The expression (\ref{orbit}) implies that the solution $u(x,y,t)$ given by
 the $\tau$-function (\ref{tau}) is a $2N$-dimensional torus orbit, $({\mathbb R}^*)^{2N}$-orbit, of a point on Gr$(N,2N)$ marked by $A_N$. Note also that
the Schubert cell $W(i_1,\ldots,i_N)$ is invariant under this torus action.
The orbits, $N$-soliton solutions, can be first classified in terms of the
Schubert decomposition with the index sets $(i_1,\ldots,i_N)$.

\begin{Example} Gr(2,4): The Schubert decomposition of Gr(2,4) is given by
\[
{\rm Gr}(2,4)=\bigsqcup_{1\le i,j \le 4}W(i,j)\,.
\]
There are six cells $W(i,j)$ with dim$\,W(i,j)=7-(i+j)$.
\begin{itemize}
\item[i)] $W(1,2)$: This is a top cell of maximum dimension four, and a point on the cell is given by
\[
A_2=\begin{pmatrix}
1 & 0 & * & * \\
0 & 1 & * & *
\end{pmatrix}\,,
\]
which includes 2-soliton solutions of T-type, $A_2^{(T)}$, and P-type, $A_2^{(P)}$. 
\item[ii)] $W(1,3)$: A point on this cell is marked by
\[
A_2=\begin{pmatrix}
1 & * & 0 & * \\
0 & 0 & 1 & *
\end{pmatrix}
\,.
\]
The 2-soliton solution of O-type with $A_2^{(O)}$ belongs to this cell. Thus
O- and P-types belong different cells, even though they show nonresonant interactions. 
\item[iii)] $W(1,4)$: The $A_2$ matrix has the form,
\[
A_2=\begin{pmatrix}
1 & * & * & 0 \\
0 & 0 & 0 & 1
\end{pmatrix}\,.
\]
In this case, nonzero minors are given by $\xi(i,4)$ for $i=1,2,3$. This implies that the exponential term $E_4=e^{\theta_4}$ becomes a common factor in the $\tau$-function. Then the solution gives $(2,1)$-soliton solution, that is,
two incoming solitons and one outgoing soliton in the asymptotics $y\to\pm\infty$
(see \cite{biondini:03}).
\item[iv)] $W(2,3)$: The $A_2$ matrix is given by
\[
A_2=\begin{pmatrix}
0 & 1 & 0 & * \\
0 & 0 & 1 & *
\end{pmatrix}\,.
\]
Then the exponential term $E_1=e^{\theta_1}$ is missing in the $\tau$-function,
and the solution describes a $(1,2)$-soliton solution.
\item[v)] $W(2,4)$: The $A_2$ matrix is
\[
A_2=\begin{pmatrix}
0 & 1 & * & 0\\
0 & 0 & 0 & 1
\end{pmatrix}\,.
\]
The solution is a 1-soliton solution of [2,3]-soliton.
\item[vi)] $W(3,4)$: The $A_2$ matrix is
\[
A_2=\begin{pmatrix}
0 & 0 & 1 & 0\\
0 & 0 & 0 & 1
\end{pmatrix}
\,.
\]
The $\tau$-function is just a product of $E_3=e^{\theta_3}$ and $E_4=e^{\theta_4}$, and the solution is trivial, $u=0$.
\end{itemize}
\end{Example}

\begin{Remark}
Since $W(1,3)$ is a boundary of the top cell $W(1,2)$,
a point on $W(1,3)$ can be obtained a limit of a sequence of points
in $W(1,2)$. For example, consider the following matrix (representing 
one parameter family of points in $W(1,2)$),
\[
A_2(\epsilon)=\begin{pmatrix}
1  &  0  & -\epsilon^{-1} & -\epsilon^{-1} \\
0  &  1  & \epsilon^{-1}+\epsilon & \epsilon^{-1}-\epsilon
\end{pmatrix}\,,
\]
where the parameter $\epsilon$ is a positive number.
If $\epsilon<1$, then the matrix gives
2-soliton solution of T-type (note that all the minors are positive).
Then in the limit $\epsilon\to 0$, the matrix can be
interpreted as the O-type matrix $A_2^{(O)}$ in $W(1,3)$, i.e.
(recall here
that $\epsilon^{-1}\tau\equiv \tau$ is the Pl\"ucker coordinate, i.e. they both give the
same solution). It is also obvious that $A_2^{(N)}$ is obtaind by
a limit of T-type matrix. Thus T-type is 
generic, while the O-type and P-type are non-generic. Also in the Gallipoli
conference in June-July 2004, Pashaev showed that 2-soliton solution of T-type is a degenerate case of 4-soliton
solution of nonresonant type (obtained by the Hirota method) \cite{pashaev:04}. 

Since $N$-soliton solution is an orbit on a particular Schubert cell
$W(i_1,\ldots,i_N)$ which is a boundary of the top cell
$W(1,\ldots,N)$, it seems to be true that all the $N$-soliton solutions are
obtained by the limits of the $\tau$-function of T-type which lives
on the top cell. We will discuss more details in a future communication.
\end{Remark}

\section{Construction of the $A_N$ matrix}
\label{Amatrix}
In this section, we give an explicit construction of the matrix $A_N$
for $N$-soliton solution parametrized by the pair $({\bf n}^+,{\bf n}^-)$:
The key idea for the construction is based on the classification of 2-soliton
solutions, that is, a local structure of the matrix contains one of those
types, O-, T- or P-types. With a given pair $({\bf n}^+,{\bf n}^-)$,
we construct an $N$-soliton solutions consisting of $N$ line solitons
with the labels $[n_j^+,n_j^-]$ for $j=1,\ldots,N$. 
In order to do this, let us first define:
\begin{Definition}
\label{length}
The length of $[n_j^+,n_j^-]$-soliton, denoted by $L_j$, is defined by
\[
L_j:={\rm min}\left\{\, n_j^--n_j^+\,,~~2N-n_j^-+n_j^+\,\right\}\,.
\]
Note that $1\le L_j\le N$ for all $j=1,\cdots,N$.
\end{Definition}
In the cases of 2-soliton solutions, we have
\begin{itemize}
\item[i)] for O-type, $L_1=L_2=1$,
\item[ii)] for T-type, $L_1=L_2=2$,
\item[iii)] for P-type, $L_1=L_2=1$.
\end{itemize}
Then we note that the length of the line soliton gives the condition on the minors. In those cases, $\xi(n_j^+,n_j^-)=0$ if and only if $L_j<2$.

For $N$-soliton solutions with $[n_j^+,n_j^-]$-solitons, we impose 
that the $\tau$-functions satisfy the following condition
which plays a crucial role for the construction of $A_N$:

\begin{Definition}\label{Nsolcondition}
 We say that a $\tau$-function 
satisfies the {\it N-soliton condition}, if it satisfies the duality (Definition \ref{duality}), and 
has missing exponential terms (i.e. vanishing minors)
if and only if there exists $[n_j^+n_j^-]$-soliton whose length is less than $N$. If $L_j<N$ for some $j$, we require that
\begin{itemize}
\item{} for $L_j=n_j^--n_j^+$, then the $N\times N$ minors satisfy
\[
\xi(\,\ldots,\,[n_j^+,n_j^++1,\ldots,n_j^-]\,,\,
\ldots\,)=0\,,
\]
\item{} for $L_j=2N-n_j^-+n_j^+$, then
\[
\xi(\,[1,\ldots,n_j^+]\,,\,\ldots\ldots\,,\,
[n_j^-,\ldots,2N]\,)=0
\]
\end{itemize}
Here the $L_j+1$ entries inside the brackets $[\cdots]$ are consecutive numbers,
and the other entries in the ``$\dots$'' outside the $[\cdots]$ can take any numbers.
\end{Definition}

Recall that in the case of $N$-soliton solutions for the Toda lattice, every soliton has the length $N$, and there is no vanishing minor \cite{biondini:03}.
The $N$-soliton condition is a natural extension of that in the case
of 2-soliton solutions, and it provides a local information of 2-soliton interactions
in $N$-soliton solution which are classified into three types, O-, T- and P-types. 
Note in particular, with this condition, one identifies all
the vanishing minors in the $\tau$-function.

Let us now construct the matrix $A_N$:
The construction has several steps, and it might be better to explain the
steps using an explict example.

\smallskip
Let us consider the case $N=4$ with ${\bf n}^+=(1,2,4,5)$ and ${\bf n}^-=(
3,7,6,8)$:

\medskip
\noindent
{\bf Step I:} Since the minors $\xi(1,2,4,5)\ne 0$ and 
$\xi(3,6,7,8)\ne 0$,  one can easily see that $A_N$ matrix has the structure,
\begin{equation}
\label{step1}
\begin{pmatrix}
*  &  *  &  *  &  0  &  0  &  0  &  0  &  0  \\
0  &  *  &  *  &  *  &  *  &  *  &  *  &  *  \\
0  &  0  &  0  &  *  &  *  &  *  &  *  &  *   \\
0  &  0  &  0  &  0  &  *  &  *  &  *  &  *   \\
\end{pmatrix}\,,
\end{equation}
where ``$*$'' marks a possible nonzero entry, and in particular,
the first element in each row should be nonzero (this is the pivot place).
Note here that $\xi(1,2,3,j)=0$ for any $j$, which is also considered
as the 4-soliton condition with the length $L_1=2$ of $[1,3]$-soliton.
Notice that $\xi(i_1,i_2,i_3,i_4)=0$ for any $4\le i_1,\ldots,i_4\le 8$
are dual to $\xi(1,2,3,j)=0$ for some $j$.

\medskip
\noindent
{\bf Step II:} In this step, we put the form (\ref{step1}) into the RREF.
First we normalize the first nonzero entries to be 1 in every rows, and eliminate the (2,4)-entry by the third row and the (3,5)-entry by the forth row. Note that this process does not
affect the 0's of the matrix in (\ref{step1}). Then we have
\begin{equation}
\nonumber
\begin{pmatrix}
1  &  *  &  *  &  0  &  0  &  0  &  0  &  0   \\
0  &  1  &  *  &  0  &  0  &  *  &  *  &  *   \\
0  &  0  &  0  &  1  &  0  &  *  &  *  &  *   \\
0  &  0  &  0  &  0  &  1  &  *  &  *  &  *  \\
\end{pmatrix}\,.
\end{equation}
Further eliminating ``$*$'' in the 2nd columns by the 2nd row, we have 
\begin{equation}
\label{step2b}
\begin{pmatrix}
1  &  0  &  *  &  0  &  0  & \#  & \#  & \# \\
0  &  1  &  *  &  0  &  0  & \#  & \#  & \# \\
0  &  0  &  0  &  1  &  0  &  *  &  *  &  * \\
0  &  0  &  0  &  0  &  1  &  *  &  *  &  *  \\
\end{pmatrix}\,,
\end{equation}
where the two rows in the $2\times 3$ submatrix with $\#$ entries are parallel
(recall $\xi(5,6,7,8)=\xi(4,6,7,8)=0$).
Also note that the signs of nonzero entries are uniquely determined by
the requirement of 
all the minors to be nonnegative, for example, the last three entries in
the first row are nonpositive, and the signs of the last three entries of the following rows change
alternatively. Up to this point, we did not use the order in ${\bf n}^-$. In another words, the form of the matrix in (\ref{step2b})
is common for any ${\bf n}^-$ with a given ${\bf n}^+$.

\begin{figure}[t!]
\epsfig{file=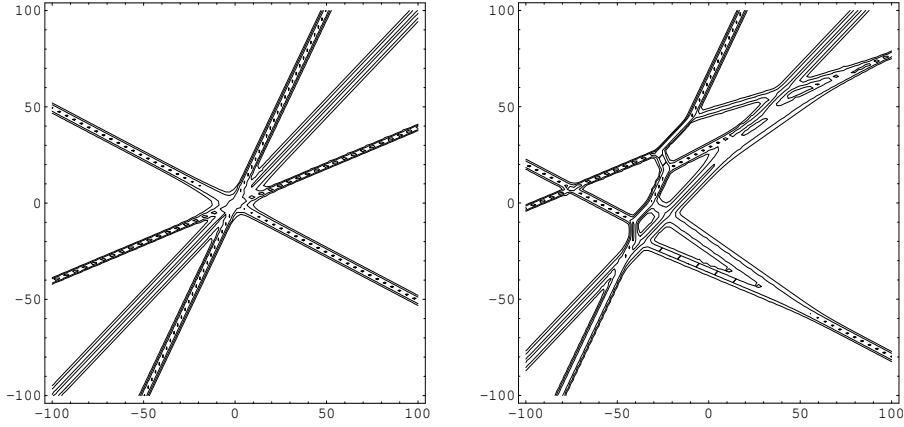,height=6cm,width=12cm}
\caption{4-soliton solutions labeled by ${\bf n}^+=(1,2,4,6)$ and
${\bf n}^-=(3,7,6,8)$. The left figure shows the 4-soliton solution
at $t=0$ and the right one at $t=10$. The parameters are given by $\theta_j^0=0$ for all $j=1,\ldots,8$, and $(k_1,\ldots,k_8)=(-3/2,-1,-1/2,0,1/2,1,3/2,2)$.
The solution at $t=-10$ is obtaind by the reflection $(-x,-y)$ of
the right graph.}
\label{4sol:fig}
\end{figure}

\medskip
\noindent
{\bf Step III:} Now we use the orders in ${\bf n}^-$ to determine an explicit
entries in the matrix for an $4$-soliton solution consisting of four line solitons labeled by $[n_j^+,n_j^-]$, i.e.
$[1,3]$, $[2,7]$, $[4,6]$, $[5,8]$. In this step, we identify all other
zero elements (i.e. zero minors) of the matrix $A_N$ using the $N$-soliton condition (Definition \ref{Nsolcondition}).
The follwoing graph shows the overlapping relations among those solitons
and the length of each soliton:
\begin{equation}
\label{4solgraph}
\quad \begin{matrix}
\circ & -    & \bullet &      &       &      &        &        &  :  & L_1=2\\
     &\circ &   -     &  -   &  -    &  -   &\bullet &        &  :  & L_2=3 \\
     &      &         &\circ & -    &\bullet&        &        &  :  & L_3=2\\
      &      &         &      & \circ &  -   &  -     &\bullet &  :  & L_4=3 \\
 \end{matrix}
\end{equation}
Now we use the 4-soliton condition. First we notice that from $L_1=2<4$
we have $\xi(1,2,3,j)=0$ for any $j$ (see Step I). Also from $L_3=2$,
we have $\xi(4,5,6,j)=0$ for any $j$. In particular, $\xi(1,4,5,6)=0$
implies that $(2,6)$-entry must be zero, so that $(1,6)$-entry also zero.
Also $\xi(3,4,5,6)=0$ gives $\xi(1,2,7,8)=0$ which is also the 4-soliton
condition with $L_2=3$. The $\xi(1,2,7,8)=0$ implies the $2\times 2$
determinant of the right bottom corner must be zero. 
We then get the following structure,
\[
\begin{pmatrix}
1  &  0  &  *  &  0  &  0  &  0  & \#      &  \#  \\
0  &  1  &  *  &  0  &  0  &  0  & \#      &  \#  \\
0  &  0  &  0  &  1  &  0  &  *  & \times  & \times  \\
0  &  0  &  0  &  0  &  1  &  *  & \times  & \times  \\
\end{pmatrix}\,,
\]
where the $2\times 2$ minor with the $\times$ entries is zero.
 Also from the partial overlaps (i.e. T-type interaction) in $[1,3]$ with $[2,7]$,
 in $[2,7]$ with $[5,8]$ and in $[4,6]$ with $[5,8]$, we need
 nonzero entries for those marked by $*$, $\#$ and $\times$.
Then one can easily find an explicit example of the matrix $A_4$
whose $4\times 4$ minors are all nonnegative, 
\begin{equation}
\nonumber
\begin{pmatrix}
1  &  0  &  -2 &  0  &  0  &  0  & -3  &  -9   \\
0  &  1  &  1  &  0  &  0  &  0  &  1  &  3   \\
0  &  0  &  0  &  1  &  0  & -1  & -2  &  -2   \\
0  &  0  &  0  &  0  &  1  &  1  &  1  &  1    \\
\end{pmatrix}\,.
\end{equation}
In Figure \ref{4sol:fig}, we illustrate a 4-soliton solution
generated with this matrix.
Notice that there are three resonant vertices corresponding to the
interactions between $[1,3]$ and $[2,7]$, $[2,7]$ and $[5,8]$,
and $[4,6]$ and $[5,8]$ (see the graph
\ref{4solgraph}). Other three vertices are two of O-types
and one of P-type.

One can show the following Proposition for the general case:
\begin{Proposition}
\label{shortL}
If there is a pair $[n_j^+,n_j^-]$ with $L_j=n_j^--n_j^+<N$,
then the $N\times (L_j+1)$ submatrix $(a_{mn})$ with ${1\le m\le N}$ and
$n_j^+\le n\le n_j^-$ of the matrix $A_N$ has the maximum rank $L_j$.
(Similar thing is also true for the case $L_j=2N-n_j^-+n_j^+$.)
\end{Proposition}
\begin{Proof}
Since $L_j<N$, we have $\xi([n_j^+,\cdots,n_j^-],*,\cdots,*)=0$
(see the $N$-soliton condition). Here the entries marked with $*$ are
arbitrary columns of $A_N$ whose maximum rank is $N$. If we
choose $N-L_j-1$ independent columns for those entries, then the $\xi=0$
implies that the submatrix can have at most the rank $L_j$.
\end{Proof}
Then the following is imediate as a Corollary of this Proposition:
\begin{Corollary}
\label{shortestL}
If a pair $[n_j^+,n_j^-]$ has the minimum length, i.e. $n_j^-=n_j^++1$,
then the $N\times 2$ submatrix with the columns $(a_{mn})$ with
$1\le m\le N$ and $n=n_j^+,\,n_j^-$ of the matrix should be in the form,
\[\left[
\begin{matrix}
0      &    0 \\
\vdots & \vdots\\
0      &    0  \\
1      &    1  \\
0     &     0  \\
\vdots  & \vdots\\
0      &   0  \\
\end{matrix}\right]\,,\quad 
{\rm where~1's~are~in~the~}j{\rm th~ row.}
\]
\end{Corollary}
\begin{Proof}
 From Proposition \ref{shortL}, the rank of the submatrix is one. This implies that the rows of the submatrix are all proportional. Since the entry
 $(j,n_j^+)$ is the pivot 1, all the other entries of this column should be
 zero in the RREF. This gives the result.
 \end{Proof}
 This Corollary shows that the line soliton marked by $[n_j^+,n_j^-]$
 of length one intersects with all other solitons without resonance, that is,
 the interactions are either O-type or P-type. Notice that in this case
 $[n_j^+,n_j^-]$ cannot have partial overlap with other pairs, and this
 can be seen in the matrix $A_N$.
Proposition \ref{shortL} provides the interaction structure of $N$-soliton
solution for O-type and P-type.

We now show that the Young diagrams $Y^+$ and $Y^-$ provide some
information on the interaction pattern of $N$-soliton solution.
If we ignore the phase shifts and resonances among the solitons, $N$-soliton solution has 
$\frac{N(N-1)}{2}$ vertices in a generic situation. Then one can find the
number of vertices having a particular type of interactions.
\begin{Theorem}
\label{num-res}
The $N$-soliton solution labeled by the pair of Young diagrams ($Y^+$,
 $Y^-$) has
\begin{itemize}
\item{}
$|Y^+|$ number of vertices of O-type, and
\item{}
$|Y^-|$ number of vertices of P-type.
\end{itemize}
The number of resonant vertices, T-type, is then given by
\[
\frac{N(N-1)}{2}-\left|Y^+\right|-\left|Y^-\right|\,.
\]
\end{Theorem}
\begin{Proof}
The Young diagram $Y^+$ can be obtained directly from the matrix $A_N$:
For example, consider ${\bf n}^+=(1,2,4,5,7)$ and ${\bf n}^-=(3,9,6,10,8)$.
Then the graph showing the overlaps among $[n_j^+,n_j^-]$ is given by
\[
\begin{matrix}
\circ  &   -   &\bullet &       &       &    &       &    &    &     \\
       & \circ & -      &    -  &   -   &   - & -   & - & \bullet  &   \\
       &       &\square &\circ  & -    &  \bullet &    &   &   &    \\
       &       &\square &      &\circ  &  -  &  -   & -  &- & \bullet   \\
       &      & \square &       &      &\square&\circ & \bullet& & \\
\end{matrix}
\]
where the boxes under the non-pivots ($\bullet$'s) give $Y^+$,
which is given by $(1,1,2)$ representing the number of boxes in the row from the bottom ($Y^+$ shape is just up-side-down picture of one given in the graph). It is then
obvious that each $\square$ provides O-type interaction, i.e. 
two solitons having no overlap. For example, $[1,3]$ has no overlap with
$[4,6]$, $[5,10]$ and $[7,8]$.
This proves that $|Y^+|$ gives the number of vertices of O-type.

The diagram $Y^-$ gives the number of {\it reverce} in the sequence
$(n_1^-,\ldots,n_N^-)$. Again consider the example above. The number $n_2^-=9$ has two reverces, 6 and 8. Then from the graph above, this implies that
$[2,9]$ has complete overlap with $[4,6]$ and $[7,8]$, i.e. $[2,9]$-soliton
has P-type interaction with those solitons. Now it is obvious that
$|Y^-|$ gives the number of vertices of P-type. 
\end{Proof}
       
\medskip
Before ending this section, we give a complete list of the 3-soliton solutions with explicit examples of $A_3$ matrices:

\smallskip
\noindent
{\bf Case a:} ${\bf n}^+=(1,2,3)$: Then we have $m_{{\bf n}^+}=1\cdot 2\cdot 3=6$
different choice of ${\bf n}^-$. that is, six different 3-soliton solutions.
 Since the Young diagram $Y^+$ has no box,
$Y^+=\emptyset$, there is no O-type vertex in those solutions.
\smallskip

\begin{itemize}
\item[a1.] ${\bf n}^-=(4,5,6)$: This is the case of Toda lattice \cite{biondini:03}. The pair of Young diagrams is $(\emptyset,\emptyset)$,
that is, all the vertices are T-type. Example of the matrix $A_3$ is given by the following matrix. We also show the overlapping graph.
\[
\begin{pmatrix}
1  &  0  &  0  &  1  &  3  &  6  \\
0  &  1  &  0  & -1  & -2  & -3  \\
0  &  0  &  1  &  1  &  1  &  1  
\end{pmatrix}\,, \quad\quad
\begin{matrix}
\circ & -    &   -   & \bullet &        &       \\
      &\circ &  -    &   -     &\bullet &       \\
      &      &\circ  &   -     &   -    & \bullet \\
 \end{matrix}
\]
\item[a2.] ${\bf n}^-=(4,6,5)$: This has two T-type vertices and one P-type
vertex, i.e. $Y^-=\square$. In particular, the P-type interaction between $[2,6]$- and $[3,5]$-
solitons implies that the (3,6) entry should be zero (because of $\xi(1,2,6)=0$). Then we have an example
\[
\begin{pmatrix}
1  &  0  &  0  &  1  &  2  &  4  \\
0  &  1  &  0  & -1  & -2  & -3  \\
0  &  0  &  1  &  1  &  1  &  0  
\end{pmatrix}\,,  \quad\quad
\begin{matrix}
\circ & -    &   -   & \bullet &        &       \\
      &\circ &  -    &   -     & -      &\bullet  \\
      &      &\circ  &   -     & \bullet&        \\
 \end{matrix}
\]
Note here that the duality of $\xi(1,2,6)=0$ and $\xi(3,4,5)=0$. Both
vanishing minors are due to the 3-soliton condition for $L_2=L_3=2$.
\item[a3.] ${\bf n}^-=(5,4,6)$: This has again two T-types vertices and one P-type vertex (notice that the Young diagram $Y^-$ is the same as the
case 2a). An example of $A_3$ is
\[
\begin{pmatrix}
1  &  0  &  0  &  0  &  1  &  2  \\
0  &  1  &  0  & -1  & -2  & -2  \\
0  &  0  &  1  &  1  &  1  &  1  
\end{pmatrix}\,,\quad\quad
\begin{matrix}
\circ & -    &   -   &  -       &\bullet &        \\
      &\circ &  -    &\bullet   &        &     \\
      &      &\circ  &   -     &   -     & \bullet  \\
 \end{matrix}
\]
Again notice the 3-soliton conditions for $L_1=L_2=2$ which give
 $\xi(1,5,6)=0$ and its dual $\xi(2,3,4)=0$. The $\xi(2,3,4)$ then gives
 $(1,4)$-entry to be zero.
\item[a4.] ${\bf n}^-=(5,6,4)$: There are two P-type and one T-type vertices.
We have
\[
\begin{pmatrix}
1  &  0  &  0  &  0  &  1  &  1  \\
0  &  1  &  0  &  0  & -2  & -1  \\
0  &  0  &  1  &  1  &  0  &  0  
\end{pmatrix}\,,\quad\quad
\begin{matrix}
\circ & -    &   -   &   -     &\bullet &        \\
      &\circ &  -    &   -     &   -    &\bullet  \\
      &      &\circ  & \bullet &        &      \\
 \end{matrix}
\]
Note here that $[3,4]$-soliton has P-type interaction with other two solitons,
which gives the last row to be $(0,0,1,1,0,0)$ and 3rd and 4th columns to be
$e_3\in{\mathbb R}^3$ (see Corollary \ref{shortestL}).
\item[a5.] ${\bf n}^-=(6,4,5)$: This case is similar to the previous one, i.e.
two P-types and one T-type. $A_3$ is
\[
\begin{pmatrix}
1  &  0  &  0  &  0  &  0  &  1  \\
0  &  1  &  0  & -1  & -2  &  0  \\
0  &  0  &  1  &  1  &  1  &  0  
\end{pmatrix}\,,\quad\quad
\begin{matrix}
\circ & -    &   -   & -       &     -  &\bullet   \\
      &\circ &  -    &\bullet  &        &      \\
      &      &\circ  &   -     & \bullet &    \\
 \end{matrix}
\]
\item[a6.] ${\bf n}^-=(6,5,4)$: All the vertices are of P-type, and the matrix
$A_3$ can be written by
\[
\begin{pmatrix}
1  &  0  &  0  &  0  &  0  &  1  \\
0  &  1  &  0  &  0  & -1  &  0  \\
0  &  0  &  1  &  1  &  0  &  0  
\end{pmatrix}\,,\quad\quad
\begin{matrix}
\circ & -    &   -   &   -     &   -    &\bullet \\
      &\circ &  -    &   -     &\bullet  &         \\
      &      &\circ  & \bullet &         &     \\
 \end{matrix}
\]
The Young diagram $Y^-$ is maximum with three boxes.
\end{itemize}

In those examples, one shoud note that the pattern of the zeros in the last
three columns (non-pivot columns) presents the shape of the 
corresponding Young diagram $Y^-$.

\medskip
\noindent
{\bf Case b:} ${\bf n}^+=(1,2,4)$: There are $m_{{\bf n}^+}=1\cdot 2\cdot 2=4$ different
types of 3-soliton solutions. Now the $Y^+$ has one box, there is one O-type
interaction in those solutions.

\smallskip
\begin{itemize}
\item[b1.] ${\bf n}^-=(3,5,6)$: There are two T-type vertices and one O-type.
An example of $A_3$ matrix is
\[
\begin{pmatrix}
1  &  0  & -1  &  0  &  1  &  2  \\
0  &  1  &  2  &  0  & -1  & -2  \\
0  &  0  &  0  &  1  &  1  &  1  
\end{pmatrix}\, ,\quad\quad
\begin{matrix}
\circ & -    &\bullet  &        &       &    \\
      &\circ &  -    &   -     &\bullet  &         \\
      &      &       & \circ   &   -     & \bullet    \\
 \end{matrix}
\]
Notice here the duality in $\xi(1,2,3)=0$ and $\xi(4,5,6)=0$.
\item[b2.] ${\bf n}^-=(3,6,5)$: The $Y^-$ has one box, and the vertices
are all different types, O-, N- and T-types. The matrix $A_3$ can be
\[
\begin{pmatrix}
1  &  0  & -1  &  0  &  0  &  2  \\
0  &  1  &  1  &  0  &  0  & -1  \\
0  &  0  &  0  &  1  &  1  &  0  
\end{pmatrix}\, ,\quad\quad
\begin{matrix}
\circ & -    &\bullet  &        &       &    \\
      &\circ &  -    &   -     &    -   &\bullet     \\
      &      &       & \circ   & \bullet  &      \\
 \end{matrix}
\]
\item[b3.] ${\bf n}^-=(5,3,6)$: This is similar to the previous case of b2.
$A_3$ can be
\[
\begin{pmatrix}
1  &  0  &  0  &  0  &  1  &  2  \\
0  &  1  &  1  &  0  &  0  &  0  \\
0  &  0  &  0  &  1  &  1  &  1  
\end{pmatrix}\, ,\quad\quad
\begin{matrix}
\circ & -    &  -    &    -    &\bullet  &        \\
      &\circ &\bullet  &       &         &       \\
      &      &       & \circ   &   -     & \bullet    \\
 \end{matrix}
\]
\item[b4.] ${\bf n}^-=(6,3,5)$: The $Y^-$ has two boxes, and there are
two P-type vertices and one O-type. The matrix $A_3$ is given by
\[
\begin{pmatrix}
1  &  0  &  0  &  0  &  0  &  1  \\
0  &  1  &  1  &  0  &  0  &  0  \\
0  &  0  &  0  &  1  &  1  &  0  
\end{pmatrix}\, ,\quad\quad
\begin{matrix}
\circ & -    &   -   &   -     &    -    &\bullet   \\
      &\circ &\bullet  &       &         &      \\
      &      &       & \circ   & \bullet  &   \\
 \end{matrix}
\]
\end{itemize}
 \begin{figure}[t!]
\epsfig{file=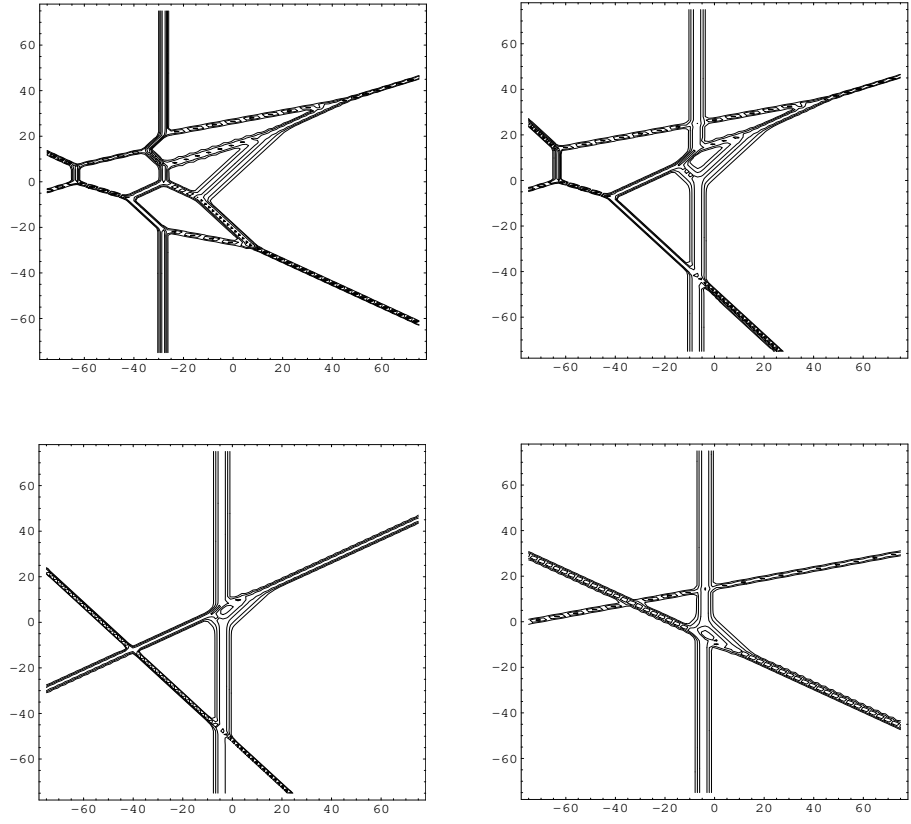,height=11cm,width=12cm}
\caption{3-soliton solutions. All the cases assume $\theta_j^0=0$ for all $j=1,\ldots,6$. The top left figure shows the solution in the case a1 with
the parameters $(k_1,\ldots,k_6)=(-3,-2,0,1,2,3)$ and $t=7$. The
top right one shows the case a3 with $(-3,-1,0,1,2,3)$ and $t=7$. The one in the bottom left shows the case a5 with
$(-4,-1,0,1,2,3)$ and $t=4$, and the one in the bottom right shows the case c1 with $(-2,-1,0,1,2,3)$ and $t=4$.}
\label{3sol:fig}
\end{figure}

\medskip
\noindent
{\bf Case c:} ${\bf n}^+=(1,2,5)$: There are $m_{{\bf n}^+}=1\cdot 2\cdot 1=2$ different
3-soliton solutions. Since the $Y^+$ has two horizontal boxes, there are
two O-type vertices in those solutions.
\smallskip

\begin{itemize}
\item[c1.] ${\bf n}^-=(3,4,6)$: The $Y^-$ has no box, and the 3-soliton solution
has one T-type interaction with two O-types. An matrix $A_3$ can be
\[
\begin{pmatrix}
1  &  0  & -1  & -2  &  0  &  0  \\
0  &  1  &  1  &  1  &  0  &  0  \\
0  &  0  &  0  &  0  &  1  &  1  
\end{pmatrix}\, ,\quad\quad
\begin{matrix}
\circ & -    &\bullet  &        &       &    \\
      &\circ &  -     &\bullet  &       &         \\
      &      &       &         & \circ   & \bullet    \\
 \end{matrix}
\]
\item[c2.] ${\bf n}^-=(4,3,6)$: The $Y^-$ has one box, and the solution has
one P-type interaction with two O-types. The matrix $A_3$ is given by
\[
\begin{pmatrix}
1  &  0  &  0  & -1  &  0  &  0  \\
0  &  1  &  1  &  0  &  0  &  0  \\
0  &  0  &  0  &  0  &  1  &  1  
\end{pmatrix}\, ,\quad\quad
\begin{matrix}
\circ & -    &   -    &\bullet  &        &      \\
      &\circ &\bullet  &       &         &    \\
      &      &       &         & \circ   & \bullet    \\
 \end{matrix}
\]
\end{itemize}

\medskip
\noindent
{\bf Case d:} ${\bf n}^+=(1,3,4)$: The $Y^+$ has two boxes, and there are $m_{{\bf n}^+}=
1\cdot 1\cdot 2=2$ different 3-soliton solutions which have two O-type
interactions.
\smallskip

\begin{itemize}
\item[d1.] ${\bf n}^-=(2,5,6)$: Since $Y^-=\emptyset$, there is one T-type
intersection with two O-types.
An example of the matrix $A_3$ is
\[
\begin{pmatrix}
1  &  1  &  0  &  0  &  0  &  0  \\
0  &  0  &  1  &  0  & -1  & -2  \\
0  &  0  &  0  &  1  &  1  &  1  
\end{pmatrix}\, ,\quad\quad
\begin{matrix}
\circ &\bullet  &        &       &      &     \\
      &         &  \circ &  -     &\bullet  &          \\
      &         &         & \circ  &   -    & \bullet    \\
 \end{matrix}
\]
\item[d2.] ${\bf n}^-=(2,6,5)$: Now $Y^-$ has one box, and the solution has
one P-type interaction with two O-types. The matrix $A_3$ is given by
\[
\begin{pmatrix}
1  &  1  &  0  &  0  &  0  &  0  \\
0  &  0  &  1  &  0  &  0  & -1  \\
0  &  0  &  0  &  1  &  1  &  0  
\end{pmatrix}\, ,\quad\quad
\begin{matrix}
\circ &\bullet  &        &       &      &     \\
      &         &  \circ &  -     &  -  &\bullet          \\
      &         &         & \circ  & \bullet &     \\
 \end{matrix}
\]
\end{itemize}

\medskip
\noindent
{\bf Case e:} ${\bf n}^+=(1,3,5)$: This is the ordinary 3-soliton solution with the matrix,
\[
\begin{pmatrix}
1  &  1  &  0  &  0  &  0  &  0  \\
0  &  0  &  1  &  1  &  0  &  0  \\
0  &  0  &  0  &  0  &  1  &  1  
\end{pmatrix}\, ,\quad\quad
\begin{matrix}
\circ &\bullet  &        &       &      &     \\
      &         &  \circ &\bullet  &       &     \\
      &         &        &        & \circ   & \bullet    \\
 \end{matrix}
\]
Some of the 3-soliton solutions are illustrated in Figure \ref{3sol:fig}.

\section{$N$-soliton solutions of the KdV equation}\label{KdV}
\begin{figure}[t!]
\epsfig{file=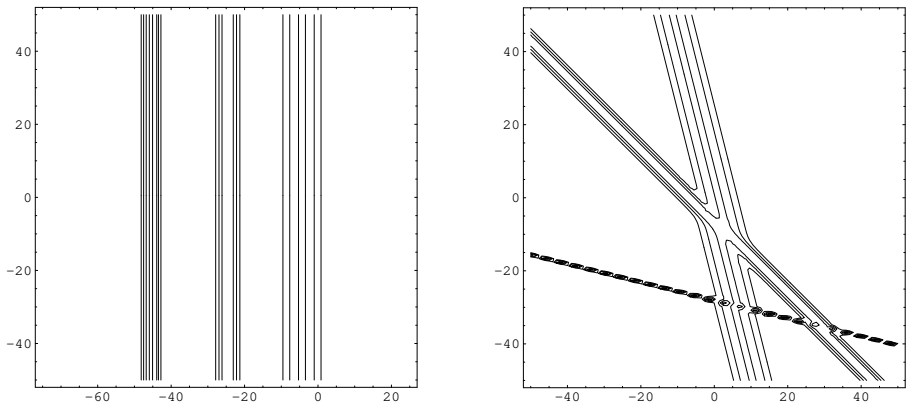,height=5.5cm,width=12cm}
\caption{3-soliton solutions of the KdV equation labeled by ${\bf n}^+=(1,2,3)$ and
${\bf n}^-=(6,5,4)$. The left figure shows the 3-soliton solution in $x$-$y$ plane
at $t=25$ with $\theta_j=0,~\forall j$. The right one shows the solution
in the $x$-$t$ plane with different $\theta_j$'s. The $k_j$'s are given by $(k_1,\ldots,k_6)=(-4/3,-1,-1/2,1/2,1,4/3)$.}
\label{3solkdv:fig}
\end{figure}

Here we consider $N$-soliton solutions of the KdV equation, and show that
they cannot have a resonant interaction.
$N$-soliton solutions of the KdV equation can be obtained by the 
constraint $\partial u/\partial y=0$ in the KP equation.
Since line solitons are given by $[n_j^+,n_j^-]$-solitons for $j=1,\ldots,N$,
the constraint implies that all the solitons are parallel to the $y$-sxis, i.e.
\[
c_j=k_{n_j^+}+k_{n_j^-}=0 \quad {\rm for~all}~~j=1,\ldots,N.
\]
 Now from the ordering $k_1<\cdots<k_{2N}$, we assume $k_j$'s to satisfy
\[
k_1<k_2<\cdots<k_N<0,\quad k_{N+j}=-k_{N-j+1}\quad {\rm for}~~j=1,\ldots,N.
\] 
Then we take the set $({\bf n}^+,{\bf n}^-)$ as
\[
n_j^+=j,\quad n_j^-=2N-j+1\quad {\rm for}~~ j=1,\ldots,N,
\]
which leads to $c_j=0$ for all $j=1,\ldots,N$. In terms of the matrix $A_N$, this corresponds to
\[
A_N=\begin{pmatrix}
1      &   0   &  \cdots  & 0      & 0    & \cdots & 0 & * \\
0      &   0   &  \cdots  & 0      & 0    & \cdots &* & 0  \\
\vdots &\vdots &\ddots    &\vdots  &\vdots& \ddots &\vdots&\vdots\\
0      & 0     &  \cdots  &  1     & *   &  \cdots& 0    &  0\\
\end{pmatrix}\,.
\]
This matrix indicates that any pair of line solitons are of P-type, that is,
all the interactions are nonresonant. Each line soliton has the following form,
\[
u(x,t)=2k_j^2\,{\rm sech}^2\,\theta_j\,,\quad {\rm with}\quad \theta_j=-k_jx-k_j^3t+\theta_j^0\,.
\]
Thus each soliton has the velocity $dx/dt=-k_j^2$. We illustrate a 3-soliton solution of the KdV equation in Figure \ref{3solkdv:fig}.

\vskip 0.5cm
\noindent
{\it Acknowledgements.}
The author would like to thank G. Biondini for valuable discussions
and a Mathematica program for the calculation of the $\tau$-functions.
He also thanks to O. Pashaev for showing him the reference \cite{pashaev:04}
before publication and for valuable discussions on AKNS reductions
of the KP equation.
This research is partially supported by the NSF grant DMS0404931.

\bibliographystyle{amsalpha}

\end{document}